\documentstyle[12pt,russian]{article}
\begin{document}
\centerline{\large Deformation of environment having string structure}
\centerline{\large in view of the topological characteristics}
\centerline{\it Trinh V.K.}
\centerline{\it Department of physics of polymers and crystals MGU}
\centerline{\it E-mail :  khoa@polly.phys.msu.ru}
\bigskip
\hangindent=2cm \hangafter =-3 \noindent
   Summary: In this paper the new model  of plastic deformation is
constructed.
      For classification of plastic
     statuses the theory of knot  is used.
\large

1. In the paper [1] the relation  between the theory of knot  and
   the statistical  mechanics are shown.
   In this clause the theory of knots  creates the new device and
the large opportunities to consider  precisely - solublely
the statistical mechanics.
   Here occurs dynamic symmetry in
the evolutionary equation.
   In the most simple case, definition of
invariant polinomial  of the knot  is equivalent to definition of function
of
distributions for two-dimantion  models in a critical point.
   In
the same  sense definition of invariant polinomial of knot
is equivalent to definition of integral Feimann  ,
means, that each diagram Feimann corresponds  invariant polinomial of
knot.
    In research of 2-dimantion  model a condition of
integrability  is connected to a subgroup of group braid as the equation
 Yang - Basters .
    In the theory Chern-Simons-Witten the statistical
 sum
  is covariantly equivalent with " Link invariant ".
    Here we use a method of
 presentation of groups  to construct the  integral Feimann  for
the  function of distribution of probability in the probem of plastic
deformation in
  the scheme of work [7].

2. Last years there  is the extensive experimental data ,
 allowing to formulate the new approaches to problems of plasticity.
      Among them it deserves  special attention to the  synergetics
approach,
 considering  a deformable firm body as open,
 tightly nonequilibrium  system, and plastic deformation - as
 dissipative  process lowering level of elastic stress of environment
 [2,3].
      In essence, in a condition of self-organizing it  is soliton  with
certain
  topological characteristics [7].
      In the present  time the wave
  character of plastic deformation of firm bodies does not cause doubts.

      According to the synergetics  approach the density of distribution
of
 probabilities of transition is the silution of
 the equation Fokker - Planck.
      However, using potential character  of the
process of deformation, instead of the equation Fokker - Planck
it is possible to use
 the equation Schrodingers .
      So, shall consider this equation as
$$
\frac{\partial f}{\partial t} = \frac{\partial^2 f}{\partial e^2} -
\frac{W(e)}{T} f \eqno(1)
$$
where $W (e) $ - free energy in a role of potential  of deformation,
$T$- temperature.
      It is possible to write the solution as integral Feimann :
$$
f(e, 0; e (t), t) = {\cal N} \int\exp\{-i\int\limits^t_0 d\tau G \}
{\cal D} e(t)
$$
$$
G = [\frac{W(e(\tau))}{T} + \frac{\dot e^2(\tau)}{2}] \eqno (2)
$$
and the point above $e $ designates derivative on time.

3. Basic our idea in such approach is, that we consider
strings as trajectories of deformation in the mechanics of
 continuous environment.
      For simplicity
we consider  only a 2-dimention case, it means that  these trajectories
 are located on a surface in the space Oclid .
      The configuration of space of system n of strings is
 $$
\begin{array}{l}
M_n = \{(Z_1, \ldots, Z_n):~~ Z_i\ne Z_j, i\ne j \}
\end{array} \eqno(3)
$$
 Topology of a configuration of space not trivial.
     It is necessary
 to investigate not equivalent homotopic integrals Feimann.
    The homotopic  class will be determined through " winding number " n.
     Let's consider two strings $a $ and $b $.
     Number n is determined for part of a curve $C $, which does not pass
    through
 string b, located in $ \xi $:
$$
\begin {array} {l}
  n = \frac{1}{2\pi i}\int\limits_C\frac {dz}{z-\xi}
\end{array} \eqno (4)
$$
     More general, if we choose coordinates with two axes, it is possible
to determine
 " winding angle " $ \theta $ curve $C$ for a string $b $
$$
\begin{array}{l}
\theta =\theta_1-\theta_o+2\pi n
\end{array} \eqno (5)
$$
where $ \theta_o $ and $ \theta_1 $ are corners of an initial and final
point
of the  curve.
      However, each string has
 internal space.
      In genera , we can choose internal
 this  space as group  Lie $G $ .
      So, it is possible to generalize " number of turns ",
 which will receive the following values in this internal space:
$$
\begin{array}{l}
\theta=sign(C) \mid\theta_1-\theta_o\mid+2\pi w \\
w=nT_a\otimes T_b =\frac{1}{2\pi i}\int\limits_C\frac {dz}{z-\xi}
T_a\otimes T_b
\end{array}\eqno(6)
$$
where $T_a, T_b $ is a presentation  of strings a and b.
      We carry out  parametrize  of a trajectory through $z (t)$,
 $t_o\le t \le t_1 $.
     So, we consider  propagatory Feimann  of a string $a $ from a
      place $z_o=z(t_o) $
 to a place
 $ z_1=z t_1) $.
    Similarly with [4] we have the propagatory Feimann  for
 homotopic    class $l $ as
 $$
\begin{array}{l}
 K_l(z_1, t_1, z_o, t_o) = \int D_l z(t)D_l\overline {z}(t)
 \exp(i\int\limits_{t_o}^{t_1}Gdt)\delta^2(2\pi lT_a\otimes T_b-\theta)
\end{array}\eqno(7)
$$
where $G(3)$  were given in the solution  of equation Schrodinger.
    Using Forier presentation, for function Dirac  we shall receive
$$
\begin{array}{l}
 \delta^2(2\pi lT_a\otimes T_b-\theta)
 = \int\frac{dkd\overline {k}}{4\pi^2}\exp[-i(k\phi +\overline {k}\phi)]
 \\
 P\exp\{i[2\pi k(lT_a\otimes T_b -w)] +c.c \}
\end{array}\eqno(8)
$$
where $c.c$ - it is the member of approximation of the high order.
     It is the  description as
functional integral of topological property of the configuration
spaces.
     As there was a self-organizing in process of
plastic deformation, has appeared not trivial topological
"interaction "
between strings.
     Substituting (8) in (7) it is received
 $$
\begin{array}{l}
 K_l(z_1, t_1, z_o, t_o) =
                        \int\frac{dkd\overline {k}}{4\pi^2}
                       \exp[-i(k\phi +\overline {k}\phi)] \\
 .\overline{K_l}(z_1,t_1,z_o,t_o; k, \overline {k}) \\
  \overline{K_l}(z_1,t_1,z_o,t_o; k, \overline {k}) =
  \int D_l z(t)D_l\overline{z}(t)
 \exp(i\int\limits_{t_o}^{t_1}Gdt) \\
\qquad .\exp \{i\int\limits_{t_o}^{t_1}[k(\frac{iz^.}{z-\xi} +
(1\pi l)T_a\otimes T_b+c.c]dt \}
\end{array}\eqno(9)
$$
    For homotopic  class $ (l_1, \ldots, l_{i-1}, l_{i+1}, \ldots, l_n) $,
designating a difference of an initial and final corner of a trajectory i
concerning trajectory j through
 $ \phi_{ij} $:
 $$
\begin{array}{l}
\phi_{ij} =sign(C_i)\mid\theta_{ij1} -\theta_{ijo} \mid
\end{array}\eqno(10)
$$
 we receive propargator  Feimann  for a the trajectory i, having
the presentation $T_i $:
 $$
\begin{array}{l}
  K_{li}(z_{1i}, t_1, z_{oi}, t_o) =
                        \int\frac{dkd\overline {k}}{4\pi^2}
                       \exp[-i\sum\limits_{i, j=1}^{n}(k\phi_{ij} +
                       \overline{k}\phi_{ij})]
                       \\
 .\overline{K}_{li}(z_{li1}, t_1, z_{lio}, t_o; k, \overline{k}) \\
  \overline{K}_{li}(z_{i1}, t_1, z_{io}, t_o; k, \overline{k}) =
  \int D_l z(t)D_l\overline{z}(t)
 \exp(i\int\limits_{t_o}^{t_1}Gdt) \\
\qquad .\exp\{i\int\limits_{t_o}^{t_1}[k\sum_{j=1, j\ne}^{n}
( \frac{iz_i^.}{z_i-z_j} +1\pi l_j)T_i\otimes
 T_j+c.c]dt \}
\end{array}\eqno(11)
$$
4. Environment  with string structure is an environment  such as polymer
or
material of composites [5].
     In a usual situation
it is considered neutral.
     Research a status of deformation of such environment
was executed on model Flory  in [5].
     In [6] it is considered
the influence of topology on deformation.
     We identify a string with an ideal polymeric circuit.
     In any
the point of a circuit,  from definition of deformation,  we  can receive
$$
< dl '^2 > - <dl^2> = \ll 2e\gg <dl^2>
$$
where $ \ll \gg $ - average value on distribution of probability of
 deformation,
$ < > $ - average value on Gauss distribution for ideal polymer.
     It means, that it is possible to receive Gauss distribution through
the distribution of
probabilities of deformation and vice versa .
     In this message the deformation was considered  as
the process of fuctuation .
     So, for each group $G $ and its  presentation there is
the statistical sum or function of distribution.
   the   topological
 character of this group plays a main role during formation of the
plastic structure.
     In the classical theory of plasticity every
 trajectory determines the appropriate plastic status.
     The interaction between trajectories was not taken into account.
     Here exists
not trivial topology because there is  interaction of strings in process
of
self-organizing.

    In a result the theory of strings of plasticity is constructed on
 following scheme.

    Let
 in beginning in
body there are elastic strings or long elastic areas of a type
 strings (for example, macromolecular of polymers [5]).
     In a status of self-organizing each string has group and its
presentation.
     For each type of interaction of strings one is determined the
statistical
 sum or the integral Feimann , o.  exists classification of
plastic status with the help of invariant  of knot.
     The probability of transition is determined due to this propagator .

\end{document}